\documentclass[11pt, a4paper]{article}
\usepackage[german, english]{babel}
\usepackage{a4wide}
\usepackage{ifthen}
\usepackage{amsmath}
\usepackage{amssymb}

\newcommand{\diag}{\operatorname{diag}}

\newcommand{\qb}{\bar{q}}
\newcommand{\q}{q}

\renewcommand{\det}{\operatorname{Det}}
\newcommand{\tr}{\operatorname{Tr}}
\newcommand{\sgn}{\operatorname{sgn}}

\newcommand{\const}{\text{const.}}

\newcommand{\Gr}[2][]{\ensuremath{
                       \mathrm{#2}
                      \ifthenelse{\equal{#1}{}} {} {(#1)} } }
\newcommand{\lie}{\mathfrak}
\newcommand{\gr}[2][]{\ensuremath{
                      \lie{#2}
                      \ifthenelse{\equal{#1}{}} {} {(#1)} } }    

\begin{document}
\author{J. Budczies and Y. Shnir\\
        {\footnotesize \it
        Universit\"at zu K\"oln, Institut f\"ur Theoretische Physik,
        Z\"ulpicher Str. 77, 50937 K\"oln, Germany.}}
\date{~}
\title{Color-Flavor Transformation for the Special Unitary 
Group and Application to Low Energy QCD} 
\maketitle

\begin{abstract}
The color-flavor transformation for the unitary group
(Zirnbauer 1996) is extended to the special unitary group.
The resulting partition function is represented as sum over 
disconnected sectors characterized by a $\Gr[1]{U}$-charge. 
Application to low energy QCD on a lattice leads to a theory
where the inverse number of colors appears as expansion parameter.  
We use a saddle point approximation to estimate the partition 
function both in the pure mesonic sector and in the case of a 
single baryon on a mesonic background.
\end{abstract} 

\section{Introduction}
Effective Lagrangians have been seen to work well in the description 
of low energy properties of mesons and hadrons, since their
introduction \cite{Weinberg:79}. Initially these Lagrangians were 
introduced {\it ad hoc} by considering the symmetries of the underlying 
theory, QCD. Later concrete calculations schemes were given to 
connect the effective action with the QCD low energy 
lattice action, see for example \cite{Kawamoto:81, Myint:94a}.
The purpose of this work is to introduce a new method for calculating 
a effective action, which rests on a recently discovered transformation 
\cite{Zirnbauer:96}. The starting point is the Euclidean $\Gr[N_c]{SU}$ 
lattice gauge theory \cite{Wilson:76} described by the action
\begin{equation}
\label{QCDaction}
  S = S_{\text{gauge}} + S_{\text{quarks}},
\end{equation}
where $S_{\text{gauge}}$ is the standard kinetic term for the Yang-Mills
action on the lattice 
\begin{equation}
\label{gaugeaction}
  S_{\text{gauge}} = - \frac{1}{4g^2} 
  \sum_{\text{plaquettes}} \tr(UUU^\dagger U^\dagger) + \text{c.c.}
\end{equation}
and
\begin{equation}
\begin{split}
  S_{\text{quarks}} = \frac{1}{2a}
  \sum_{\text{sites } n} &\sum_{\mu=1}^d \left(
   \qb^i_a(n) \gamma_\mu U^{\dagger{ij}}_\mu(n) \q^j_a (n+e_\mu) - 
  \qb^i_a(n+e_\mu) \gamma_\mu U_\mu^{ij}(n) \q^j_a(n) \right) \\
   &+\qb^i_a(n) M_{ab} \q^i_b(n).     
\end{split}
\end{equation}
is a discrete counterpart of the Dirac operator, which couples the 
quarks to the gauge field. Throughout this work we consider the strong 
coupling limit and neglect the action $S_{\text{gauge}}$.

In this brief report we will give only results, while proofs and 
details of the calculations will be presented elsewhere. 
The paper is organized as follows:
we start by illustrating the color flavor transformation for 
the cases of $\Gr[N_c]{U}$ and $\Gr[N_c]{SU}$ (sect.~2). 
In sect.~3 we apply this technique to transform the partition 
function of strong coupling QCD and arrive at a decomposition of the 
partition function into disconnected sectors characterized by 
the baryonic charge. The integration over the quarks is performed in
the pure mesonic sector and leads to a theory of a collective 
field $Z$, with $N_c$ appearing as a factor in front of the action
(sect.~4). A saddle point approximation is used to estimate the
partition function in the leading order of the inverse number 
of colors (sect.~5). We make use of a long distance approximation 
in combination with a gradient expansion to recover the corresponding
low energy effective action. In our approach the lattice constant $a$
is a parameter, which has to be fixed at a certain value to yield 
to realistic values of the observables. In sect.~6 we give first results 
for the case of a single baryon on a mesonic background.

\section{Color-Flavor Transformation}

We start with a simple example, which illustrates the general
form of the transformation \cite{Zirnbauer:98}.
For four fermion fields $\qb_+, \q_+$ and $\qb_-, \q_-$ one can 
easily check the relation
\begin{equation}
\begin{split}
\label{u11}
  &\frac{1}{2 \pi} \int_0^{2 \pi} du \; 
   \exp( \qb_+ e^{iu} \q_+ + \qb_- e^{-iu} \q_-) = 1 +\qb_+ \q_+ \qb_- \q_-
  = \\
  &2 \int_\mathbb{C} \frac{d\mu(z,\bar{z})}{1+z \bar{z}} \;
  \exp(\qb_+ z \qb_- - \q_+ \bar{z} \q_-).
\end{split}
\end{equation}
The integration measure
$d \mu(z, \bar{z}) = \frac{\const}{(1+z \bar{z})^2}\, dz d\bar{z}$
is the natural measure of two-sphere $S^2$ in stereographical coordinates. 
In ref. \cite{Zirnbauer:96} the following remarkable generalization
of the above scheme was given:
\begin{equation}
\begin{split}
\label{uncnf}
   &\int_{U(N_c)} dU \;
    \exp(\qb_{+a}^i U^{ij} \q_{+a}^j + \qb_{-b}^i \bar{U}^{ij} \q_{-b}^j)
   = \\ 
   &\int_{\mathbb{C}^{N_f \times N_f}} 
    \frac{D \mu(Z, Z^\dagger)}{\det(1+ZZ^\dagger)^{N_c}} \;
    \exp \left(\qb_{+a}^i Z_{ab} \qb_{-b}^i -
               \q_{+a}^i \bar{Z}_{ab} \q_{-b}^i \right).
\end{split}
\end{equation}
The fermions are now vectors of $N_f$ "flavors" ($a,b = 1,...,N_f$) 
and $N_c$ "colors" ($i,j = 1,..., N_c$). 
The field $Z$ becomes a $N_f \times N_f$-matrix and the integration
measure
\begin{equation}
\label{measure}
  D \mu(Z, Z^\dagger) = \frac{\const}{\det(1+ZZ^\dagger)^{2N_f}} \, 
                        dZ dZ^\dagger
\end{equation}
is the invariant measure on the compact coset space
$\Gr[2 N_f]{U} / \Gr[N_f]{U} \times \Gr[N_f]{U}$.
For obvious reasons equation \eqref{uncnf} is called the color-flavor
transformation for the color group $\Gr[N_c]{U}$.
Note that for $N_f=1$ the coset space is isomorphic to the two-sphere
($\Gr[2]{U} / \Gr[1]{U} \times \Gr[1]{U} \cong S^2$). 
Thus for $N_f=1$ the general scheme reduces to the special case
considered before.

There are color-flavor transformations for several other groups 
\cite{Zirnbauer:98, Beuchelt:97}.
Having an application to QCD in mind, we consider the special unitary
group $\Gr[N_c]{SU}$, which is the gauge group of the strong
interaction ($N_c=3$). 
Again we start with the simplest case $N_c=1$ and $N_f=1$, where the 
integral on the l.h.s. of \eqref{u11} reduces to a single point
(evaluation at unity), and the following statement is immediate: 
\begin{equation}
\begin{split} 
  &\exp( \qb_+ \q_+ + \qb_- \q_-) = 
  1 + \qb_+ \q_+ + \qb_- \q_- + \qb_+ \q_+ \qb_- \q_- = \\
  &2 \int_\mathbb{C} \frac{d\mu(z,\bar{z})}{1+z \bar{z}} \;
  \exp(\qb_+ z \qb_- - \q_+ \bar{z} \q_-)
  (1 + \frac{1}{2} \qb_+ (1+z \bar{z}) \q_+ 
     + \frac{1}{2}\qb_- (1+z \bar{z}) \q_-).
\end{split}
\end{equation}
Again, cf. \cite{BNS:00}, this is a special case of a more general scheme:
\begin{equation}
\begin{split}
\label{suncnf}
  &\int_{SU(N_c)} dU \;
  \exp(\qb_{+a}^i U^{ij} \q_{+a}^j + \qb_{-b}^i \bar{U}^{ij} \q_{-b}^j)
  = \\ 
  &\int_{\mathbb{C}^{N_f \times N_f}} 
  \frac{D \mu(Z, Z^\dagger)}{\det(1+ZZ^\dagger)^{N_c}} \;
  \exp \left(\qb_{+a}^i Z_{ab} \qb_{-b}^i -
             \q_{+a}^i \bar{Z}_{ab} \q_{-b}^i \right)
  \sum_{Q=-N_f}^{N_f} \chi_Q(Z, Z^\dagger, \q, \qb).   
\end{split}
\end{equation}
The physical interpretation of equation \eqref{suncnf}
is as follows:
It is a sum over contributions, which come from mesonic excitations
over the vacuum ($Q=0$) or mesonic excitations in the presence of 
$|Q|$ baryons ($Q>0$) or $|Q|$ antibaryons ($Q<0$).
The explicit expressions for $\chi_Q$ in the vacuum sector and
in the one-baryon sector are \cite{BNS:00}
\begin{equation}
\begin{split}
   \chi_0 = \const, \quad
   \chi_1 = \const \, \sum_{\sigma \in S_{N_c}}
              \sgn \sigma \prod_{i=1}^{N_c}
              \qb^i_{+a} (1 + Z Z^\dagger)_{ab} \q^{\sigma(i)}_{+b}.
\end{split}
\end{equation}
The sum runs over all permutations $S_{N_c}$ of the numbers
$1, ..., N_c$.

\section{Lattice Gauge Theory}
We consider a Euclidean $\Gr[N_c]{SU}$ gauge theory in $d$ dimensions 
on a hypercubic lattice with lattice constant $a$. We restrict
our considerations to an even number $d$ of spacetime dimensions. 
The fermions $\psi(n)$ are placed on the lattice sites labeled by $n$,
while the gauge field $U_\mu(n)$ is put on the lattice links,
which may be labeled by a lattice site $n$ and a direction 
$\mu = 1,..,d$.
The gauge theory is described by the partition function
\begin{equation}
  \mathcal{Z}(\q, \qb) =
  \prod_{n,\mu} \int_{\Gr[N_c]{SU}} d U_\mu(n) \, 
  \exp(-S_{n,\mu} \left(U_\mu(n)) \right),
\end{equation}
where the fermions on two neighboring sites are coupled through
the gauge field on the connecting link in the gauge invariant way
\begin{equation}
\begin{split}
\label{sulattice}
  S_{n, \mu} \left(U_\mu(n) \right) = &\frac{1}{2a} \left(
  \qb^i_a(n) \gamma_\mu U^{\dagger{ij}}_\mu(n) \q^j_a (n+e_\mu) - 
  \qb^i_a(n+e_\mu) \gamma_\mu U_\mu^{ij}(n) \q^j_a(n) \right) \\
  &+\qb^i_a(n) M_{ab} \q^i_b(n).      
\end{split} 
\end{equation}
The action \eqref{sulattice} is a discrete version of the gauge
covariant Dirac operator. Here we do not worry about the fermion
doubling problem and employ naive fermions. 
The mass matrix $M= \diag(m_1,...,m_{N_f})$
contains the quark masses and is diagonal in flavor space.
The Dirac spinor fields carry spacetime indices, which take values 
$\nu=1,...,N_s$ with $N_s=2^{d/2}$, and are coupled through the 
$\gamma$-matrices. The $\gamma$-matrices
are the generators of a Clifford algebra over the Euclidean spacetime 
and satisfy the anticommutation relations
$\gamma_\mu \gamma_\nu + \gamma_\nu \gamma_\mu = 2 \delta_{\mu \nu}$.
Now we apply the color-flavor transformation to the gauge field
$U_\mu(n)$ separately on each link. In doing so, we get a new action,
which is a function of the flavor field $Z_\mu(n)$, which is again 
situated at the links of the lattice.
The result is a sum over contributions from sectors with different 
baryonic charge $Q$ and can be written as
\begin{equation}
  \mathcal{Z}(\q, \qb) =
  \prod_{n,\mu} \int_{\mathbb{C}^{N_f \times N_f}} 
  D\mu_{N_c}(Z_\mu(n), Z_\mu^\dagger(n)) \sum_{Q=-N_f}^{N_f} 
  \chi_{Q,\mu}(n)  
  \exp \left(- S_{n, \mu}(Z_{n, \mu}) \right) 
\end{equation}
with the integration measure \eqref{measure}.
The color-flavor transformed action for a link $(n, \mu)$ reads 
\begin{equation}
\begin{split}
  S_{n, \mu}(Z_\mu(n)) = & \frac{1}{2a}
  \left(\qb_a^i(n + e_\mu) \gamma_\mu Z_{\mu, ab}(n) \q_b^i(n+e_\mu) +
        \qb_a^i(n) \gamma_\mu Z_{\mu,ab}^\dagger(n) \q_b^i(n) \right) \\
  &+\qb^i_a(n) M_{ab} \q^i_b(n).
\end{split}
\end{equation}
Explicit expressions for the prefactors in the case of the vacuum sector 
($Q=0$) and the one-baryon sector ($Q=1$) are
\begin{align}
  \chi_{0, \mu}(n) &= \const, \\
  \label{chi1}
  \chi_{1, \mu}(n) &= \const 
   \sum_{\sigma \in S_{N_c}} \sgn \sigma \prod_{i=1}^{N_c}
   \qb^i_a(n+e_\mu) \gamma_\mu 
   (1 + Z_\mu Z_\mu^\dagger)_{ab}(n) \q^{\sigma(i)}_b(n).
\end{align}

\section{Integration over the Quarks and Saddle Point}
Performing the Gaussian integral over the quarks in the pure mesonic 
sector ($Q=0$) and sending the result back to the exponent, we obtain 
the action
\begin{equation}
\begin{split}
\label{zaction}
  S_{Q=0}(Z) = N_c \Biggl( -& \sum_n \tr \ln \left( \sum_{\mu=1}^d 
                \gamma_\mu \left( Z_\mu(n-e_\mu) + Z^\dagger_\mu(n) \right)
                - a M \right) \\
                +& \sum_n \sum_{\mu=1}^d \tr 
                \ln \left(1 + Z_\mu(n) Z_\mu^\dagger(n) \right) \Biggr). 
\end{split}
\end{equation}
Variation with respect to the independent variables $Z$ and $Z^\dagger$
yields the saddle point equations
\begin{equation} 
\begin{split}
   \gamma_\mu \left( \frac{1}{Z_\mu(n)} + Z^\dagger_\mu(n) \right) &= 
   \sum_{\nu=1}^d  \gamma_\nu \left(Z_\nu(n-e_\nu) + 
                             Z^\dagger_\nu(n) \right) - 2aM, \\
   \gamma_\mu \left( \frac{1}{Z^\dagger_\mu(n-e_\mu)} 
   + Z_\mu(n-e_\mu) \right) &=
   \sum_{\nu=1}^d \gamma_\nu \left(Z_\nu(n-e_\nu) + 
                             Z^\dagger_\nu(n) \right) - 2aM.
   \label{saddlenew}
\end{split}
\end{equation}
The solution of the saddle point equations is
\begin{equation}
\label{saddlepoint}
   Z_\mu^0(n) =  Z_\mu^{0 \dagger}(n)=z \gamma_\mu I \qquad
   \text{ with } 
   z = \frac{1}{2d-1} \left( aM \pm \sqrt{2d-1+ (aM)^2} \right),
\end{equation}
where $I$ is the unit matrix in flavor space.
Recall the following definitions of $\gamma_5$ and the projectors
on the chiral components of the spinors   
\begin{align}
  \gamma_5 = i^{d(d-1)/2} \gamma_1 \cdot ... \cdot \gamma_d, \\
  \gamma_L = \frac{1}{2} (1 + \gamma_5), \quad
  \gamma_R = \frac{1}{2} (1 - \gamma_5).
\end{align}
In the chiral limit ($M=0$) there is a saddle point manifold
\begin{equation}
\label{saddlemanifold}
   Z_\mu^0(n) =  Z_\mu^{0 \dagger}(n)= z \gamma_\mu g^{\gamma_5} :=
   z \gamma_\mu(\gamma_L \otimes g + \gamma_R \otimes g^{-1}),
\end{equation} 
which is parameterized by unitary matrices $g \in \text{U}(N_f)$.
The lattice gauge theory in the chiral limit ($M=0$) is invariant 
under the chiral transformations 
($(g_L, g_R) \in \Gr[N_f]{U}_L \times \Gr[N_f]{U}_R$) 
of the quarks fields
\begin{equation}
\begin{split}
   \q(n) &\rightarrow 
   \left( \gamma_L \otimes g_L + \gamma_R \otimes g_R \right) \q(n), \\
   \qb(n) &\rightarrow 
   \left( \gamma_L \otimes g_R^{-1} + 
          \gamma_R \otimes g_L^{-1} \right) \qb(n).
\end{split}
\end{equation}
The chiral symmetry of the action is preserved by the
color-flavor transformation and leads to the invariance of
the action \eqref{zaction} and the saddle point manifold
under the transformations  
\begin{equation}
\begin{split}
\label{zsymmetry}
  Z_\mu(n) &\rightarrow 
  \left(\gamma_L \otimes g_L^{-1} + \gamma_R \otimes g_R^{-1} \right)
  Z_\mu(n) \left(\gamma_L \otimes g_L + \gamma_R \otimes g_R \right),\\
  Z^\dagger_\mu(n) &\rightarrow 
  \left(\gamma_L \otimes g_L^{-1} + \gamma_R \otimes g_R^{-1} \right)
  Z^\dagger_\mu(n) 
  \left(\gamma_L \otimes g_L + \gamma_R \otimes g_R \right). 
\end{split}
\end{equation}    
The saddle point \eqref{saddlepoint} is invariant under 
\eqref{zsymmetry} only if $g_L = g_R$, i.e. it breaks the
the chiral $\Gr[N_f]{U}_L \times \Gr[N_f]{U}_R$ symmetry
to the subgroup $\Gr[N_f]{U}$. The above symmetry considerations
explain how chiral symmetry breaking takes place in our approach 
to the gauge theory. 
The chiral symmetry breaking gives rise to Goldstone
bosons, which can be identified with the coset space
$\Gr[N_f]{U}_L \times \Gr[N_f]{U}_R/\Gr[N_f]{U} \cong \Gr[N_f]{U}$
and parameterize the saddle point manifold \eqref{saddlemanifold}.  

\section{Gradient Expansion}
Our aim is to derive an effective theory, which describes the
long range behavior of Goldstone modes $g \in \Gr[N_f]{U}$.
We use the technique of a long distance approximation in
combination with a gradient expansion around the 
saddle point manifold, as it was developed in 
\cite{Altland:98, Altland:99}.
The lattice action \eqref{zaction} is replaced by a simple 
continuum action
\begin{equation}
   S_{Q=0}(Z) \longrightarrow 
   S(g) = S_{\text{fl}}(g) + S_M(g),
\end{equation} 
where $S_{\text{fl}}(g)$ is the action corresponding to 
fluctuations of the Goldstone modes, and $S_M(g)$ the
contribution due to finite quark masses.
More explicitly, $Z_\mu(n)$ is put into correspondence with a 
continuous field $g(x) \in \Gr[N_f]{U}$ in the following way:
\begin{equation}
   Z_\mu(n) = Z_\mu^{\dagger}(n)
            = z \gamma_\mu \; g(n + \frac{a}{2} e_\mu)^{\gamma_5}.
\end{equation}
Inserting into \eqref{zaction} the Taylor expansion 
\begin{equation}
   g(n+ \frac{a}{2}e_\mu) = g(n) + \frac{a}{2} \partial_\mu g(n) +
   \frac{1}{2!} \left( \frac{a}{2} \right)^2 \partial_\mu^2 g(n) + \cdots,
\end{equation}
we obtain for the hypercubic lattice
\begin{equation}
\label{actionfluc}
   S_{\text{fl}}(g) = \frac{N_s}{8d} a^{2-d} \int d^d x \; 
                       \tr \left(\partial g \partial g^{-1} \right), 
\end{equation}
\begin{equation}
\label{actionM}
   S_M(g) = N_s \frac{\sqrt{2d-1}}{2d} a^{-d}
            \int d^d x \; \tr \left(aM(g+g^{-1})\right).
\end{equation}
We have carried out the expansion up to fourth order derivatives.
In particular, we have done so for a body-centered hypercubic (BHC) 
lattice in $d=4$ spacetime dimensions. The 4-dimensional BHC lattice 
has the advantage of leading to a continuum action with complete
rotational $\Gr[4]{O}$ symmetry. 
The results will be presented elsewhere \cite{BS:00}. 

In $d=4$ spacetime dimensions both parts of the actions \eqref{actionfluc} 
and \eqref{actionM} diverge when the lattice constant approaches zero. 
Recall that we are considering the strong coupling limit only, i.e. 
we neglect the kinetic term \eqref{gaugeaction}, which is suppressed as 
$1/g^2$. Therefore our theory is restricted to the low energy sector and 
we have to keep the lattice constant at a finite value.
There are three ways to estimate the lattice constant $a$ through 
comparisons with experimental data: Using the relation
\begin{equation}
     <\qb_f \q_f> 
   = - \frac{1}{V} \frac{\partial}{\partial m_f} \bigg{|}_0 S_{\text{saddle}} 
   = - N_s N_c \frac{d-1}{d \sqrt{2d-1}} a^{1-d},
\end{equation}  
we get from the experimental value for the chiral condensate 
$a = (166 \text{ MeV})^{-1}$.
By looking at the coefficient in front of the fluctuation action,
\begin{equation}
   \frac{F_\pi^2}{4} = N_s N_c \frac{1}{8d} \, a^{2-d},
\end{equation}
we get from the experimental value for the pion decay constant 
$F_\pi$ the estimate $a = (76 \text{ MeV})^{-1}$.
By looking at the coefficient in front of the fluctuation action,
\begin{equation}
   \frac{1}{4} F_\pi^2 m_\pi^2 = 
    N_s N_c \frac{\sqrt{2 d -1}}{2 d} \, a^{1-d} m_f,
\end{equation}
we get from the experimental values for the pion decay constant, 
the mass of the pion $m_\pi$ and the mass of the light quarks $m_f$ 
the value $a = (126 \text{ MeV})^{-1}$.
We conclude that the lattice constant has to be chosen as
$a \approx (100 \text{ MeV})^{-1}  \approx 2 \text{ Fermi}$
to get a realistic description.

\section{Static Baryon}
We consider a single static baryon on a mesonic background.
In doing so we deal with the $Q=1$ sector right on the worldline 
of the baryon and the $Q=0$ sector away from the 
baryon. Performing the integration over the quarks we have to take 
into account factors in front of the exponential along the worldline 
of the baryon, which stem from the expression for $\chi_1$ \eqref{chi1}.
We introduce periodic boundary conditions in the time direction.
Introducing the quantity
\begin{equation}
   M(n) := \sum_{\mu=1}^d 
           \gamma_\mu \left( Z_\mu(n-e_\mu) + Z^\dagger_\mu(n) \right)
            - a M 
\end{equation}
we are able to express the action for the static baryon on the
mesonic background by means of the following propagator along 
the wordline
\begin{equation}
   \mathsf{G} := \prod_{n \in \text{ worldline}} G(n), \quad \text{ where } 
   G(n) := \left(1+Z_4(n)Z_4^\dagger(n)\right) M(n)^{-1} \gamma_4.
\end{equation}
The result is
\begin{equation}
\label{actionbaryon}
   S_{Q=1} = S_{Q=0} 
   -\frac{1}{N_c} \ln \sum_{\sigma \in \hat{S}_{N_c}} \mathcal{N}(\sigma)
   \prod_{l=1}^{N_c} \left( \tr \mathsf{G}^l \right)^{c_l(\sigma)} 
   + \const,
\end{equation}
where the sum extends over conjugation classes of permutations. The 
class of a permutation is determined by the lengths of its cycles. 
We denote the number of cycles of length $l$ of the permutation 
$\sigma$ by $c_l(\sigma)$. 
The normalization constants in \eqref{actionbaryon} are given by
$\mathcal{N}(\sigma) = \left( \prod_{l=1}^{N_c} l^{c_l(\sigma)} 
c_l(\sigma)! \right)^{-1}$.
For the lowest numbers of colors we get the explicit expressions 
\begin{equation}
\begin{split}
   & \tr \mathsf{G} \quad (N_c=1),\quad 
    (\tr \mathsf{G})^2 +  \tr \mathsf{G}^2 \quad (N_c=2),\\
   &(\tr \mathsf{G})^3 + 3\tr \mathsf{G}^2 \tr \mathsf{G} 
               + 2 \tr \mathsf{G}^3 \quad (N_c=3).
\end{split}
\end{equation}
The next step would be to obtain the saddle point equations
for the static baryon. In \cite{BNS:00} we study a solution 
of the saddle point equations in $d=2$ spacetime dimensions.

\section{Summary and Acknowledgments}
We have briefly described a new approach to lattice QCD,
which is useful in the low energy domain, where the quarks
are confined to mesons and baryons. 
In principle models away from the strong coupling limit
can be treated with the color flavor transformation,
but one has to replace the standard action 
\eqref{gaugeaction} by an action that can be generated with
help of an auxiliary field \cite{Zirnbauer:96, BZ:00}.

We are indebted to M. R. Zirnbauer for suggesting the subject
to us and we thank him for fruitful discussions. 
We thank S. Nonnenmacher for collaborating with us on the 
implementation of the $\Gr[N_c]{SU}$ color-flavor 
transformation and its application to the static baryon.
Useful discussions with A. Altland are acknowledged. 

\bibliographystyle{my2}
\bibliography{literature}

\begin{thebibliography}{10}
\expandafter\ifx\csname url\endcsname\relax
  \def\url#1{\texttt{#1}}\fi
\expandafter\ifx\csname urlprefix\endcsname\relax\def\urlprefix{URL }\fi
\providecommand{\eprint}[2][]{\url{#2}}

\bibitem{Weinberg:79}
S.~Weinberg, Physica \textbf{96A} (1979) 327.

\bibitem{Kawamoto:81}
N.~Kawamoto and J.~Smit, Nucl. Phys. \textbf{B192} (1981) 100.

\bibitem{Myint:94a}
S.~Myint and C.~Rebbi, Nucl. Phys. \textbf{B421} (1994) 241, hep-lat/9401009.

\bibitem{Zirnbauer:96}
M.~R. Zirnbauer, J. Phys. \textbf{A29} (1996) 7113, chao-dyn/9609007.

\bibitem{Wilson:76}
K.~Wilson, Phys. Rep. \textbf{C12} (1976) 331.

\bibitem{Zirnbauer:98}
M.~R. Zirnbauer, in \emph{Proceedings of the XIIth International Congress of
  Mathematical Physics (Brisbane, July 1997)} (1998), chao-dyn/9810016.

\bibitem{Beuchelt:97}
G.~Beuchelt, \emph{Dirac Fermions in 2+1 Dimensions with Random Mass
  Distribution}, Master's thesis, K\"oln (1997).

\bibitem{BNS:00}
J.~Budczies, S.~Nonnenmacher and Y.~Shnir, to be published.

\bibitem{Altland:98}
A.~Altland and B.~D. Simons, J.Phys. \textbf{A32} (1999) L353,
  cond-mat/9811134.

\bibitem{Altland:99}
A.~Altland and B.~D. Simons, Nucl. Phys. \textbf{B562} (1999) 445,
  cond-mat/9909152.

\bibitem{BS:00}
J.~Budczies and Y.~Shnir, to be published.

\bibitem{BZ:00}
J.~Budczies and M.~R. Zirnbauer, in preparation.

\end{thebibliography}

\end{document}